\documentclass[conference]{IEEEtran}
\IEEEoverridecommandlockouts
\usepackage{cite}
\usepackage{amsmath,amssymb,amsfonts}
\usepackage{algorithmic}
\usepackage{graphicx}
\usepackage{textcomp}
\usepackage{xcolor}
\usepackage{booktabs}
\usepackage{enumitem}
\usepackage[ruled,vlined]{algorithm2e}
\def\BibTeX{{\rm B\kern-.05em{\sc i\kern-.025em b}\kern-.08em
    T\kern-.1667em\lower.7ex\hbox{E}\kern-.125emX}}

\usepackage{xspace}

\begin{document}

\title{Privis: Towards Content-Aware Secure \\Volumetric Video Delivery}
\author{\IEEEauthorblockN{Kaiyuan Hu\textsuperscript{1}, Hong Kang\textsuperscript{1}, Yili Jin\textsuperscript{1},
Junhua Liu\textsuperscript{3},
Chengming Hu\textsuperscript{1}, Haolun Wu\textsuperscript{1}, Xue Liu*\textsuperscript{1,2}}
\IEEEauthorblockA{\textsuperscript{1}\textit{McGill University}, \textsuperscript{2}\textit{Mohamed bin Zayed University of Artificial Intelligence},\textsuperscript{3}\textit{University of Southern California}}
\IEEEauthorblockA{\textit {\{kaiyuan.hu, hong.kang, yili.jin, chengming.hu, haolun.wu\}@mail.mcgill.ca, junhua.liu.0@usc.edu, xueliu@cs.mcgill.ca}}
}

\maketitle

\begin{abstract}
Volumetric video has emerged as a key paradigm in eXtended Reality (XR) and immersive multimedia because it enables highly interactive, spatially consistent 3D experiences. However, the transport-layer security for such 3D content remains largely unaddressed. Existing volumetric streaming pipelines inherit uniform encryption schemes from 2D video, overlooking the heterogeneous privacy sensitivity of different geometry and the strict motion-to-photon latency constraints of real-time XR.

We take an initial step toward \textit{content-aware secure volumetric video delivery} by introducing \textit{Privis}, a saliency-guided transport framework that (i) partitions volumetric assets into independent units, (ii) applies lightweight authenticated encryption with adaptive key rotation, and (iii) employs selective traffic shaping to balance confidentiality and low latency. Privis specifies a generalized transport-layer security architecture for volumetric media, defining core abstractions and adaptive protection mechanisms. We further explore a prototype implementation and present initial latency measurements to illustrate feasibility and design tradeoffs, providing early empirical guidance toward future work on real-time, saliency-conditioned secure delivery.

\end{abstract}

\begin{IEEEkeywords}
volumetric video, privacy, multimedia communication
\end{IEEEkeywords}

\section{Introduction}

Volumetric video has emerged as a key paradigm in immersive media services, enabling users to experience interactive virtual environments with lifelike spatial presence and 6-DoF freedom. Unlike conventional 2D video, volumetric video streams explicit 3D geometry with texture or implicit neural representations, allowing natural motion parallax, viewpoint changes, and real-world-scale telepresence. This unique feature makes volumetric media fundamental to emerging XR applications, from remote collaboration~\cite{jin2023capture} and immersive communication to medical training~\cite{wu2024teleor} and cultural heritage preservation\cite{wang2024chang}. However, these benefits come at the cost of much higher bandwidth demands and tight motion-to-photon (MTP) latency budgets, as feedback must remain consistent with user movement to maintain comfort and presence.

While recent advances in compression and adaptive delivery have improved scalability, transport security for volumetric media remains largely unexplored. Existing solutions inherit uniform end-to-end encryption from 2D streaming pipelines \cite{tang2020vvsec}, treating all content equally and relying on uniform end-to-end encryption while assuming reliable delivery. However, this breaks down in 6-DoF settings: retransmissions and head-of-line blocking violates real-time MTP budgets, while moving security handling to application layer complicates deployment.

Beyond latency, volumetric streams pose amplified privacy risks. They reveal detailed biometric and behavioral cues. For example, facial geometry, body motion, hand gestures, and interaction context can enable identity inference, profiling, or environment reconstruction~\cite{miller2020personal, jarrold2013social, loucks2019you, werner2009use, tarnanas2013ecological}. When streamed via cloud services or content delivery networks, intermediate nodes may inspect encrypted traffic patterns, allowing leakage through packet size, rate, or inter-arrival correlations. These side channels are particularly problematic in XR, where traffic patterns correlate with user behaviors and scene semantics.

These observations expose two transport-level challenges for volumetric delivery: 
(i) tight MTP budgets are easily violated by reliable delivery and per-frame cryptographic work, and 
(ii) even with payload encryption, packet size/timing still enables traffic inference about scene dynamics and user activity. 
We therefore advocate \textbf{content-aware secure volumetric transport} that adapts protection strength to joint perceptual–privacy saliency in 3D content.

In this work, we present \textbf{Privis}, a content-aware transport framework that (i) partitions volumetric data into independent protection units, (ii) applies lightweight authenticated encryption with adaptive key rotation, and (iii) performs selective traffic shaping. Privis prioritizes privacy-critical 3D regions and user-centric geometry, while applying efficient protection to low-saliency background structures to preserve latency.

In summary, this paper makes the following contributions:

\begin{itemize}[leftmargin=*]
    \item We propose the first \textbf{transport-layer framework for secure volumetric video delivery}, enabling content-aware protection that adapts to the heterogeneous sensitivity of 3D data.

    \item We design a \textbf{perceptual-privacy saliency} model and a cube-level adaptive pipeline combining multi-level authenticated encryption, dynamic keying, and selective traffic shaping.
    
    \item We explore the feasibility of applying \textbf{saliency-conditioned, cube-wise secure delivery} within volumetric streaming pipelines, demonstrating its potential through an initial implementation and latency measurements.
\end{itemize}

\begin{figure*}[t]
\centering
\includegraphics[width=\textwidth]{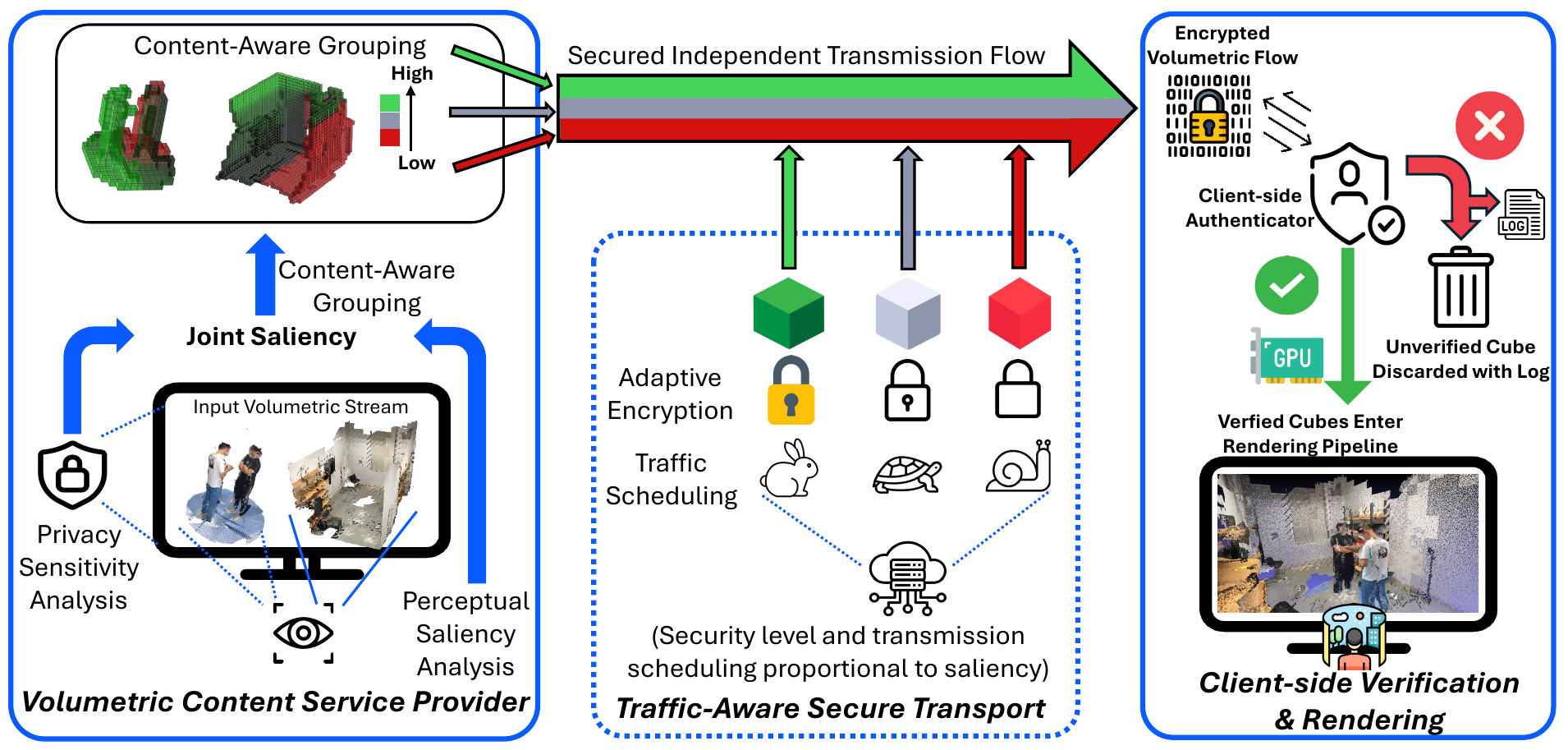}
\caption{\textit{Privis} performs saliency-driven cube partitioning, adaptive encryption, and latency-aware transmission for volumetric video. High-saliency regions are prioritized and strongly protected, while low-saliency content uses lighter security to maintain real-time performance. Client-side verification enforces decrypt-before-render, ensuring integrity in immersive delivery.
}
\label{fig:privis_pipeline}
\end{figure*}

\section{Background and Motivation}

\subsection{Volumetric Video Streaming}
Volumetric video streaming has become a key technology for immersive communication, enabling real-time transmission of 3D representations such as point clouds, meshes, or Gaussian splat. Early efforts in this field primarily focused on efficient compression and scalable delivery to address the extreme data volume of 3D content. MPEG has standardized formats such as Point Cloud Compression and Video-Based Point Cloud Compression (V-PCC)~\cite{graziosi2020overview}, while current prototypes explored adaptive streaming based on user navigation or predicted viewport trajectories~\cite{hu2023understanding}. They aim to reduce bitrate while maintaining visual fidelity and interactivity.

Recent advances explored multi-layer or foveated rendering pipelines that exploit viewer gaze or region-of-interest (ROI) to allocate bandwidth efficiently~\cite{liu2023cav3,hu2025livevv}. These systems leverage perceptual saliency in immersive delivery: they allocate resources to high-attention regions while relaxing quality in lower salience areas, improving rendering and streaming performance with negligible impact on perceived visual quality.

Beyond purely viewport-driven foveation, hybrid pipelines have emerged that combine explicit 3D geometry with neural scene representations to reduce the transmission and rendering costs. For instance, Pagoda \cite{lu2023pagoda}, a multi-layer neural radiance field streaming framework that dynamically allocates updates across different layers to balance quality and compute, which highlights the potential of scene structure and spatio-temporal dynamics to guide adaptive transmission; however, these systems assume trusted networks and do not consider security as a objective. In contrast, our work aligns protection with perceptual and privacy saliency under real-time constraints.

Meanwhile, current privacy-preserving methods for immersive video streaming focus on application-layer obfuscation or gaze-assisted masking~\cite{jin2024privacy}. While effective for 2D viewport protection, their safeguards operate on image projections rather than on explicit 3D geometry, leaving raw volumetric content vulnerable during transport. As a result, existing immersive pipelines lack confidentiality and integrity guarantees for volumetric streams—particularly against traffic analysis or partial compromise, creating a significant gap between perceptual efficiency and secure, high-fidelity 6-DoF delivery.

\subsection{Security and Privacy in Immersive Media}
\noindent\textbf{Privacy Risks in Immersive Content.} 
Immersive and volumetric applications raise distinct privacy challenges as both 3D content and interaction data reveal sensitive personal information \cite{hu2025securetelemed}. Biometric cues (e.g., facial geometry, body shape, motion patterns) and behavioral cues (e.g., gaze trajectory, gestures) can enable identity inference, profiling, or health-state estimation and are encoded by 3D representations. Unlike 2D video, these modalities continuously expose fine-grained spatiotemporal data, amplifying privacy risk.

\noindent\textbf{Limitations of Existing Solutions.} 
Prior studies on immersive privacy primarily rely on \textit{visual obfuscation} or \textit{gaze-assisted masking} at the application layer~\cite{hu2025securetelemed}, and \textit{offline encryption} for static 3D assets~\cite{giaretta2024security}. However, these methods neither offer real-time, content-aware protection nor address traffic-level leakage during transmission. These limitations motivate our \textit{Privis} that applies saliency-conditioned encryption and adaptive traffic shaping directly to volumetric video delivery.

\section{Design of Privis Framework}
In this section, we propose a generalized framework, \textit{Privis} for securing the transport of volumetric data. Rather than proposing a new codec or rendering scheme, Privis outlines an architectural direction for integrating 
content-awareness and adaptive protection into existing volumetric streaming pipelines, enabling the potential to guide future implementations toward transport-layer mechanisms that can scale with the 
heterogeneity and privacy sensitivity of 3D content.

\subsection{System Overview}

The framework overview is shown in Figure \ref{fig:privis_pipeline}. It operates as a transport layer sitting between the volumetric encoder and the network transport stack. 
At a high level, \textit{Privis} performs three operations: 
(1) analyzing 3D content to estimate joint perceptual-privacy saliency for spatial regions,
(2) applying selective protection that adapts encryption frequency, scope, and metadata to handle those regions, and 
(3) managing secure delivery while balancing real-time constraints. 

This approach shifts the focus from uniform encryption toward context-dependent protection: 
high-sensitivity 3D structures receive stronger or more frequent protection, while background regions 
are treated with lighter mechanisms to preserve latency and bandwidth.

\subsection{Content-Aware Grouping}

Explicit volumetric content consists of heterogeneous geometric and semantic elements, where different regions of a scene 
carry vastly different levels of privacy sensitivity. For example, faces, hands, and body geometry may reveal identity and behavioral 
traits, while background structures often contain little personal information. Therefore, Privis introduces content-aware 
grouping as the foundation of its security design. To guide adaptive protection, Privis assigns each spatial cube $c$ a joint saliency
score $s(c)$ that reflects both perceptual importance and privacy sensitivity:

\vspace{-5pt}

\begin{equation}
s(c) = \alpha \, \Phi_{p}(c) + (1-\alpha)\, \Phi_{s}(c), \qquad s(c)\in[0,1]
\label{eq:saliency_score}
\end{equation}

where $\alpha \in [0,1]$ balances the perceptual fidelity and
privacy protection. Higher values of $s(c)$ indicate cubes that require stronger
cryptographic protection and traffic analysis defenses.

We formalize the perceptual and privacy terms as:
\begin{equation}
\begin{aligned}
\Phi_{p}(c) &= f\!\big(\text{density}(c),\, \text{motion}(c),\, \text{view\_proximity}(c)\big), \\
\Phi_{s}(c) &= g\!\big(\text{identity\_exposure}(c),\, \text{user\_proximity}(c)\big).
\end{aligned}
\end{equation}

where $\Phi_{p}(c)$ estimates a cube's contribution to perceptual fidelity, while $\Phi_{s}(c)$ captures privacy sensitivity associated with personally revealing content. The specific cues used for $\Phi_{s}$ and $\Phi_{p}$ are not restricted to faces, hands, or body geometry; these are illustrative examples for telepresence. In practice, sensitive content varies across applications. For instance, patient anatomy in medical tele-surgery, object locations in industrial environments, or contextual background in domestic telepresence. Privis therefore treats $\Phi_{p}$ and $\Phi_{s}$ as flexible modules: both $f(\cdot)$ and $g(\cdot)$ may be analytic functions or learned models suited to the deployment context. Higher joint saliency leads to more frequent key rotation and stricter traffic shaping, while low-saliency cubes receive lightweight protection to preserve latency.

This formulation generalizes across volumetric representations and supports learned or rule–based predictors, enabling Privis to adapt to diverse application contexts such as telepresence, XR collaboration, or medical imaging.

Intuitively, regions near a user's face, hands, or interaction zone may be salient due to both visual relevance and privacy risk, while background geometry receives lower saliency. Unlike 2D streaming pipelines where saliency focuses on perceptual compression, Privis treats saliency as a security signal, enabling fine-grained and privacy-aware transport decisions.

In general, each volumetric frame is decomposed into a set of spatial cubes, each cube representing a localized 3D region that serves as the unit of encryption, transmission, and integrity verification. These cubes form the basic secure transport units in Privis and provide a consistent abstraction for applying adaptive protection
across spatially distinct parts of the scene. Each cube’s sensitivity can be estimated from multiple cues, including its spatial locality, perceptual importance, and temporal dynamics. For instance, cubes near human faces or hands are assigned higher protection priority, while static background cubes receive lightweight encryption and longer key lifetimes. This saliency-aware partitioning enables Privis to align protection cost with privacy risk while maintaining real-time throughput.

While saliency provides the guiding signal, Privis does not enforce a fixed cube formation method. Cubes may be generated via voxelization, spatial clustering, or application-level semantic cues. The key requirement is that each cube functions as an independent transport and protection unit, allowing the system to flexibly adapt security to content sensitivity
and streaming constraints.

\textbf{Temporal Cube Reuse.}
Volumetric content typically exhibits strong temporal coherence across consecutive frames. To avoid recomputing cube assignments at every frame, Privis reuses cube boundaries by default, updating them only when significant scene motion or structural change is detected. This temporal grouping amortizes saliency estimation and cube partitioning across short frame windows, reducing per-frame computation while preserving fine-grained protection. By exploiting temporal stability in 3D scenes, Privis maintains real-time responsiveness without sacrificing adaptive security control. Critically, this modular structure supports compartmentalization, failures or tampering in one region do not propagate globally, and graceful degradation, where dropped low-saliency cubes do not interrupt interactive streaming. Together, these properties enable scalable, privacy-adaptive volumetric delivery under tight latency budgets.

\vspace{15pt}

\begin{algorithm}[!ht]
\caption{Content-Aware Secure VV Delivery}
\label{alg:privis}
\KwIn{Volumetric frame $F$, saliency model $\Phi$, leakage budget $\epsilon$}
\KwOut{Protected transmission streams $\{T(c)\}$}
\vspace{15pt}
\SetKwBlock{Stage}{Stage}{end}
\Stage{Content-aware cube segmentation and scoring}{
    $C \leftarrow \text{Partition}(F)$ \tcp*{Spatial cubes}
    \ForEach{cube $c \in C$}{
        $s(c) \leftarrow \Phi(c)$ \tcp*{Joint perceptual and privacy saliency}
    }
    $C \leftarrow \text{Sort}_{desc}(C, s(c))$
}

\Stage{Adaptive key allocation}{
    \ForEach{cube $c \in C$}{
        $p(c) \leftarrow \text{ProtectionLevel}(s(c))$ \tcp*{Level: low/med/high}
        \uIf{$\text{TemporalStable}(c)$}{
            $k(c) \leftarrow \text{ReuseKey}(c)$
        }
        \Else{
            $k(c) \leftarrow \text{DeriveKey}(k_{root}, c)$
        }
    }
}

\Stage{Per-cube encryption and traffic shaping}{
    \ForEach{cube $c \in C$}{
        $payload(c) \leftarrow \text{Encrypt}(c, k(c), p(c))$

        \uIf{$s(c) > \theta$}{
            $\ell' \leftarrow \ell + \delta(s(c))$ \tcp*{Saliency-proportional padding}
            $t' \leftarrow t + \eta(s(c))$ \tcp*{Jitter injection}
            $\Delta t \leftarrow \max(\Delta t, \tau(s(c)))$ \tcp*{Guard window scheduling}
        }
        \Else{
            $(\ell', t') \leftarrow (\ell, t)$ \tcp*{No shaping for low-saliency cubes}
        }

        $T(c) \leftarrow \text{Transmit}(payload(c), \ell', t')$
    }
}

\Stage{Leakage bound check}{
    \If{$I(s(c); T(c)) > \epsilon$}{
        $\theta \leftarrow \theta - \Delta$ \tcp*{Tighten threshold}
        \text{Reassign shaping and re-transmit critical cubes}
    }
}
\end{algorithm}


\subsection{Adaptive Protection and Key Management}

Privis allocates cryptographic resources proportionally to the sensitivity of each volumetric cube, avoiding the inefficiency of uniform encryption. Instead, the system adapts protection granularity to balance confidentiality, computation, and real-time delivery requirements, ensuring privacy-critical regions receive stronger guarantees without compromising MTP latency.
\vspace{0.1in}
\vspace{0.1in}

\textbf{Protection Levels.}
Each cube is assigned a security policy specifying its encryption scope, key update cadence, and metadata-handling strategy. Higher-saliency cubes receive broader payload coverage, more frequent re-keying, and tighter traffic obfuscation, whereas low-saliency regions adopt lightweight policies to preserve responsiveness.

\textbf{Key Lifecycle.}
Privis derives cube-level keys from a session root key, refreshing them according to saliency and temporal stability. When cube boundaries remain consistent across frames, keys are reused to reduce re-key overhead while maintaining forward secrecy and bounding exposure risk.

\textbf{Formalization.}
Let $s(c) \in [0,1]$ denote the joint saliency of cube $c$. Privis maps saliency to a protection tuple:
\begin{equation}
p(c) = \mathcal{P}(s(c))
= \big(k(c),\, \rho(c),\, \sigma(c)\big),
\label{eq:protection_mapping}
\end{equation}
where $k(c)$ is key-update rate, $\rho(c)$ is encryption scope (e.g., geometry only vs. full payload), and $\sigma(c)$ encodes traffic-shaping strength.

Protection increases monotonically with saliency under latency constraints:
\begin{equation}
\begin{aligned}
\text{if } s(c_1) > s(c_2) &\Rightarrow \mathcal{P}(s(c_1)) \succeq \mathcal{P}(s(c_2)), \\
\text{s.t. } &\sum_{c} \text{Cost}(\mathcal{P}(s(c))) \le \Gamma,
\end{aligned}
\label{eq:latency_constraint}
\end{equation}

where $\Gamma$ is the real-time latency budget, ensuring proportional protection while respecting  MTP limits.

\textbf{Summary.}
This design enforces (i) stronger safeguards for privacy-critical content, (ii) efficient reuse of keys when geometry is stable, and (iii) saliency-driven shaping to mitigate inference risk. The resulting pipeline provides fine-grained confidentiality and integrity aligned to immersive-media latency constraints.

\subsection{Traffic-Aware Secure Transport}

Privis enforces transport–layer confidentiality and leakage resilience by decoupling volumetric content into independent flows and shaping traffic
according to content saliency. This design prevents adversaries from 
inferring sensitive spatial regions through observable packet patterns 
while preserving real-time delivery.

\textbf{Saliency-Driven Independent Flows.}
Privis assigns each spatial cube to an independent transport flow whose
priority, key rotation rate, and encryption scope follow content saliency.
By decoupling cubes at transport level, Privis enables: (i) parallel
encryption and transmission to reduce queue buildup, (ii) localized loss
isolation to avoid scene-wide stalls, and (iii) saliency-aware scheduling
that prioritizes privacy-critical or perceptually essential regions under
network stress. This granularity provides a flexible abstraction compatible
with low-latency protocols such as QUIC and SRTP.

\textbf{Traffic Awareness and Leakage Reduction.}
Beyond payload encryption, Privis incorporates traffic–level defenses to mitigate inference attacks that exploit packet timing, size, or transmission burst patterns. Let each cube $c \in \mathcal{C}$ have saliency $s(c)$ and assigned protection level $p(c)$; the transport objective is to prevent adversaries from inferring $s(c)$ from observable traffic statistics.

We denote the observable traffic trace for cube $c$ as
\[
T(c) = \{(\ell_i, t_i)\}_{i=1}^{N(c)},
\]
where $\ell_i$ and $t_i$ denote packet size and inter–arrival time. A leakage function $\mathcal{L}$ captures the mutual information between saliency and traffic:
\[
\mathcal{L} = I(S;T) = I(s(c); T(c)).
\]
A secure transport aims to bound $\mathcal{L}$ such that
\[
I(s(c);T(c)) \le \epsilon,
\]
where $\epsilon$ is the tolerable leakage budget.

To achieve this, Privis applies selective traffic shaping on high–saliency cubes, including:

\vspace{-0.2em}
\begin{itemize}[leftmargin=*]
\item \textit{Saliency–dependent padding:}
\[
\ell_{i}^{\prime} = \ell_{i} + \delta(c), \qquad 
\delta(c) \sim \mathcal{D}(s(c)),
\]
where $\mathcal{D}_{\text{pad}}(s)$ is a saliency-proportional noise distribution with bounded support to respect the MTP budget.

\item \textit{Jitter randomization:}\[
t_{i}^{\prime} = t_{i} + \eta(c), \qquad 
\eta(c)\sim \mathcal{J}(s(c)),
\]
which decorrelates packet timing from content dynamics while keeping delay within budget.
\item \textit{Burst scheduling with guard windows:}
\[
\Delta t(c) \ge \tau(s(c)),
\]
enforcing saliency-aware pacing and guard windows to mask short-term activity bursts.
\end{itemize}

To defend against traffic–inference attacks that attempt to
recover spatial saliency from packet timing and size patterns,
Privis applies shaping only where needed rather than padding
the entire stream. For cubes with $s(c) > \theta$, selective
padding, jitter, and pacing introduce uncertainty into the
traffic trace, constraining mutual information $I(s(c);T(c))$
within the leakage budget~$\epsilon$.

By activating shaping only on high-saliency cubes, Privis
avoids unnecessary delay on benign traffic while ensuring that
sensitive regions are indistinguishable from one another under
traffic analysis. Operating at cube granularity also prevents
shaping in a particular region from propagating latency to the
whole scene, preserving real-time volumetric updates and
interactive head-motion responsiveness.

\subsection{Client-Side Verification and Rendering Admission}

At the client, Privis enforces a strict decrypt-before-render boundary: all cryptographic processing completes before any volumetric data enters the rendering pipeline. This design minimizes the plaintext attack surface and prevents exposure of sensitive 3D content within GPU buffers, shader memory, or client-side caching paths.

\textbf{Integrity Verification and Drop Policy.}
Each received cube undergoes authentication and integrity checks prior to being admitted into the rendering pipeline. When verification fails, Privis applies a conservative policy:

\begin{itemize}
\item Discarding unverified cubes to avoid rendering corrupted or adversarial 3D content;
\item Reusing the last verified cube when current data fails authentication, 
preserving continuity without expensive real-time repair or smoothing;
\item Recording verification failures for session-level anomaly detection and replay protection.
\end{itemize}

This approach prioritizes correctness and privacy over visual completeness, reflecting the real-time and safety-critical nature of immersive communication.

\textbf{Robust Partial Rendering.}
On client side, Privis continues rendering verified cubes once they are authenticated, while isolating unverified regions. If a cube fails integrity checks, the system reuses the last trusted version, preserving motion continuity without introducing latency spikes or geometric artifacts. This per-cube hold-over policy prevents scene-wide stalls and maintains user embodiment under loss or tampering, complementing the selective transport strategy above.

\textbf{Client Trust Boundary.}
Privis assumes a trusted client execution environment and focuses protection on in-transit data. All decryption and authentication occur before volumetric content enters GPU pipelines, minimizing plaintext exposure. While techniques such as TEE-based isolation or client attestation may further harden the endpoint, they are orthogonal to our transport-layer focus and remain future integration paths.

\section{Evaluation}
To assess the effectiveness and robustness of Privis, we implemented an end-to-end prototype that integrates saliency-aware cube partitioning, adaptive key rotation, and per-cube authenticated encryption into a volumetric delivery pipeline. Our goal is to quantify the practical overhead introduced by content-aware security under realistic volumetric streaming settings and evaluate whether Privis can operate within latency budgets compatible with interactive telepresence and immersive communication. 

We focus on two key questions: (i) \textbf{Can fine-grained cube-level protection be executed in real-time?} (ii) \textbf{How does saliency-aware security compare to conventional uniform encryption?} Thus, we present cryptographic processing cost, per-frame latency, and end-to-end delay across baseline, uniform, and Privis configurations in the following parts.


\subsection{Experimental Setup}

\textbf{Prototype.}
We extend a volumetric sender–receiver pipeline with per–cube AEAD (AES–GCM) and saliency-conditioned key refresh. Cube keys are rotated on high-saliency regions and reused on stable low-saliency regions.

\textbf{Dataset and Hardware.}
We evaluate on the human-centric full-scene FSVVD dataset~\cite{hu2023fsvvd} using a server–client XR telepresence setup (RTX-class GPU at client, A100 server). Network RTT is emulated as 15\,ms.

\textbf{Cube Configuration.}
Frames are segmented into $K{=}32$–$96$ spatial cubes. Face/upper-body cubes are marked high-saliency; background cubes low-saliency, based on the dataset default setup. High-saliency cubes rekey every frame, low-saliency every $N{=}6$ frames.

\subsection{Evaluation Configurations}







We compare three configurations: \textbf{Baseline (NoEnc)}, which performs raw volumetric streaming with no protection; \textbf{Uniform-Enc}, which applies full-frame AEAD encryption to every frame; and \textbf{Privis-Adaptive}, which uses cube-level AEAD where high-saliency cubes are encrypted each frame while low-saliency cubes are refreshed every $N{=}6$ frames. This spectrum isolates the impact of content-aware adaptive protection against conventional uniform security and ideal no-encryption performance.

\subsection{Latency Metrics}

We measure the average processing latency per frame, including cube saliency scoring and grouping, key derivation or reuse, cube-level AEAD encryption, cube-level AEAD decryption, and end-to-end frame streaming time.

\subsection{Results}

Table~\ref{tab:latency1} reports mean per-frame latency (ms). We measure four representative scene in FSVVD, 1200 frames in total \cite{hu2023fsvvd}.

\begin{table}[h]
\centering
\caption{Latency Breakdown for a Single Volumetric Frame (ms)}
\label{tab:latency1}
\begin{tabular}{lccc}
\toprule
Component & NoEnc & UniformEnc & Privis-Adaptive \\
\midrule
Saliency + Grouping & 49.62 & 49.82 & 49.62 \\
Key Management & 0.01 & 0.06 & 0.04 \\
Encryption & 0.03 & 4.06 & 0.13 \\
Decryption & 0.00 & 0.08 & 0.07 \\
Transport Assembly & 0.02 & 0.01 & 0.01 \\
\midrule
\textbf{Total} & \textbf{49.88} & \textbf{58.08} & \textbf{52.98} \\
\bottomrule
\end{tabular}
\end{table}
\noindent\textbf{Note.} Saliency and grouping appear in all configurations, including NoEnc, because region partitioning is inherent to viewport-adaptive volumetric streaming.

\textbf{Findings.}
Privis introduces negligible per-frame latency overhead relative to the NoEnc baseline while remaining substantially more efficient than uniform encryption, reducing total latency by $\sim$5.1\,ms. Temporal cube reuse avoids redundant saliency computation on stable regions, and selective per-cube key refresh significantly lowers cryptographic workload while still securing privacy-critical areas.

Even with a software-only implementation, Privis remains well within interactive streaming bounds ($<20$\,ms added latency). These results confirm that fine-grained, content-aware protection preserves responsiveness while offering stronger confidentiality granularity than uniform encryption. The prototype further demonstrates feasibility without codec modification, indicating that future volumetric media pipelines can adopt saliency-guided secure transport with minimal disruption.

\section{Conclusion}

Privis shows that content-aware security can be integrated into
volumetric streaming with minimal overhead. Cube-level AEAD
and selective key refresh preserve interactive performance while substantially reducing cryptographic cost compared to uniform encryption, providing finer confidentiality without disrupting the pipeline.

Our prototype uses rule-based saliency and a constrained set of
scenes primarily to isolate Privis’s core transport-layer effects. While sufficient for demonstrating feasibility, broader evaluation across diverse content, adversarial conditions, and heterogeneous networks represents a natural next step rather than a limitation of the design. Future extensions include learned saliency models, GPU-accelerated cryptography, cross-scene adaptation, and integration with emerging volumetric codecs.

\bibliographystyle{IEEEtran}
\bibliography{ref}

\end{document}